\documentclass[aps,prl,twocolumn,showpacs,showkeys,footinbib,amsmath,amssymb,superscriptaddress]{revtex4-1}

\usepackage{amsmath}
\usepackage{amssymb}
\usepackage{graphicx}
\usepackage{bm}
\usepackage{color}
\usepackage{relsize}
\usepackage{braket}
\usepackage{bbold}
\usepackage{txfonts}
\usepackage{epstopdf}
\usepackage{hyperref}

\begin{document}

\title{Antiferroelectric Altermagnets: Antiferroelectricity Alters Magnets}
\author{Xunkai Duan}
\affiliation{Eastern Institute for Advanced Study, Eastern Institute of Technology, Ningbo, Zhejiang 315200, China}
\affiliation{School of Physics and Astronomy, Shanghai Jiao Tong University, Shanghai 200240, China}

\author{Jiayong Zhang}
\affiliation{Eastern Institute for Advanced Study, Eastern Institute of Technology, Ningbo, Zhejiang 315200, China}
\affiliation{School of Physical Science and Technology, Suzhou University of Science and Technology, Suzhou, 215009, China}
\affiliation{International Center for Quantum Design of Functional Materials (ICQD), and Hefei National Laboratory, University of Science and Technology of China, Hefei, 230026, China}

\author{Ziye Zhu}
\affiliation{Eastern Institute for Advanced Study, Eastern Institute of Technology, Ningbo, Zhejiang 315200, China}
\affiliation{International Center for Quantum Design of Functional Materials (ICQD), and Hefei National Laboratory, University of Science and Technology of China, Hefei, 230026, China}

\author{Yuntian Liu}
\affiliation{Department of Physics, University at Buffalo, State University of New York, Buffalo, New York 14260, USA}

\author{Zhenyu Zhang}
\affiliation{International Center for Quantum Design of Functional Materials (ICQD), and Hefei National Laboratory, University of Science and Technology of China, Hefei, 230026, China}
\author{Igor \v{Z}uti\'c}
\affiliation{Department of Physics, University at Buffalo, State University of New York, Buffalo, New York 14260, USA}

\author{Tong Zhou}
\email{tzhou@eitech.edu.cn}
\affiliation{Eastern Institute for Advanced Study, Eastern Institute of Technology, Ningbo, Zhejiang 315200, China}
\date{\today}

\begin{abstract}
Magnetoelectric coupling is crucial for uncovering fundamental phenomena and advancing technologies in high-density data storage and energy-efficient devices. The emergence of altermagnets, which unify the advantages of ferromagnets and antiferromagnets, offers unprecedented opportunities for magnetoelectric coupling. However, electrically tuning altermagnets remains an outstanding challenge. Here, we demonstrate how this challenge can be overcome by using antiferroelectricity and ferroelectricity to modulate the spin splitting in altermagnets, employing a universal, symmetry-based design principle supported by an effective model. We introduce an unexplored class of multiferroics: antiferroelectric altermagnets (AFEAM), where antiferroelectricity and altermagnetism coexist in a single material. From first-principles calculations, we validate the feasibility of AFEAM in well-established van der Waals metal thio(seleno)phosphates and perovskite oxides. We reveal the design of AFEAM ranging from two-dimensional monolayers to three-dimensional bulk structures. Remarkably, even a weak electric field can effectively toggle spin polarization in the AFEAM by switching between antiferroelectric and ferroelectric states. Our findings not only enrich the understanding of magnetoelectric coupling but also pave the way for electrically controlled spintronic and multiferroic devices.
\end{abstract}
\maketitle

Rather than applying magnetic field, electric control is pursued to achieve faster and more energy-efficient manipulation of magnetism~\cite{zutic2004:RMP,Tsymbal:2019,Yang2022:N}. 
The nonvolatility of magnetic materials enables not only advances in high-density storage and neuromorphic computing~\cite{zutic2004:RMP,Tsymbal:2019,Yang2022:N}, but also integrating electronics, spintronics, and 
photonics~\cite{Dainone2024:N}. Multiferroics, with their magnetoelectric coupling and both (anti)ferroelectricity and magnetism, are excellent candidates for electrical control of magnetism and spin polarization~
\cite{eerenstein2006multiferroic,Fiebig2016:evolution,Dong2019:Magnetoelectricity}. This coupling could 
also elucidate fundamental phenomena, from probing dark matter to topological phase transitions~\cite{Roising2021:PRR,chen2024ferroelectricity}.

\begin{figure}[t]
\includegraphics*[width=0.47\textwidth]{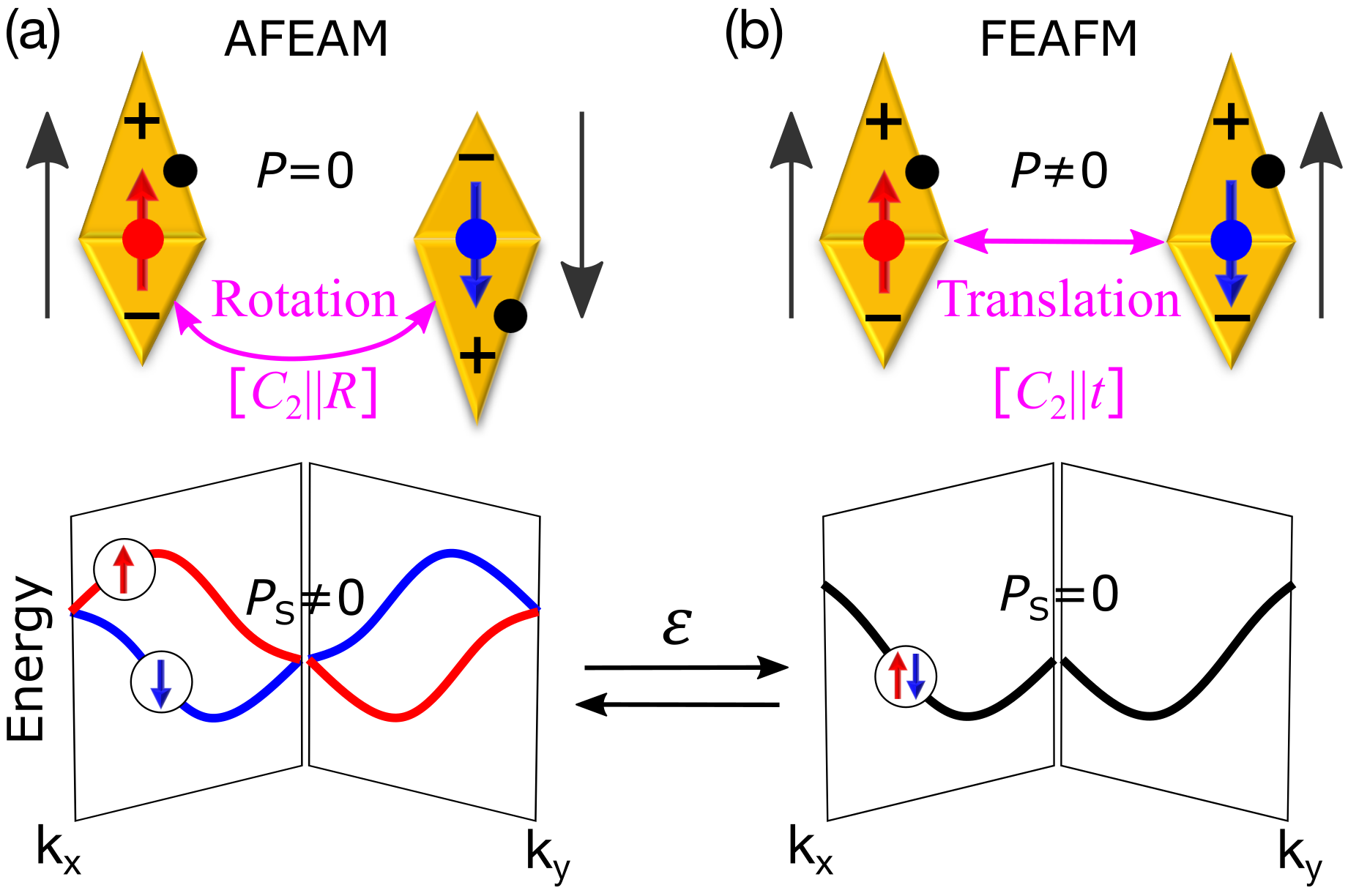}
\caption{The design principle for AFEAM and FEAFM. Magnetic atoms with opposite spins (red/blue arrows) constructing AFM sublattices. Anti-parallel (parallel) local electric polarizations (black arrows), from the asymmetric charge distributions around these atoms due to displaced nonmagnetic atoms (black dots), show AFE (FE) states without (with) net electric polarization, $\textbf{\emph{P}}$. (a) AFEAM: magnetic sublattices are connected by rotation, $R$, resulting in AM with spin polarization, $\textbf{\emph{P}}_S\neq0$. (b) FEAFM: magnetic sublattices are directly connected by translation, $t$, leading to a conventional AFM with $\textbf{\emph{P}}_S=0$. The AFEAM and FEAFM can be switched by an electric field, $\varepsilon$, enabling an electric control of $\textbf{\emph{P}}_S$.}
\label{fig:F1}
\end{figure}

Since common ferromagnets are metallic, while ferroelectric (FE) states are insulating, their coexistence in a single material is difficult~\cite{Hill2000:few}. 
Realizing multiferroics with typically insulating FE and antiferromagnets (AFM) is more feasible, while also supporting higher critical temperatures, no stray  
fields, and ultrafast dynamics, desirable for spintronics~\cite{Baltz2018:RMP,Vsmejkal2018:Nat_Phys,Chen2024:AM,Shao2024:npjspintronics}. 
However, the absence of spin polarization in conventional AFM limits their utility in multiferroics. 

Building on the understanding of AFM with the nonrelativistic spin 
splitting~\cite{Pekar1964:JETP,Litvin1974:P,Yuan2020:PRB,Hayami2020:PRB,JunweiliuNC2021,Coey:2024}, the discovery of altermagnets (AM)~\cite{Vsmejkal2022:emerging,Mazin2024:altermagnetism} with zero net magnetization and momentum-dependent spin splitting~\cite{Krempasky2024:MnTe,Osumi2024:PRB_MnTe,Lee2024:PRL_MnTe,Reimers2024:NC_CrSb,ding2024large:CrSb,zeng2024observation:CrSb,yang2024three:CrSb}, 
could overcome these challenges. The resulting phenomena with AM already include generating spin-polarized currents ~\cite{Gonzalez2021:efficient}, 
giant magnetoresistance ~\cite{Vsmejkal2022:Giant,Cui2023:Giant,samanta2024:tunneling}, nontrivial superconductivity~\cite{zhongboyanPRB2023,Brekke2023:model,AMMajorana_PRL2024}, anomalous Hall effect~\cite{PRB2023CaCrO3,Betancourt2023:spontaneous,Song2024crystalAM,Guo2023:quantum}, and other interesting spintronic behavior~\cite{shao2023neel,zhang2024:predictable,Denisov2024AM,Yan2024PRL_Skyrmion}.
Despite that AM unify many advantages of AFM and ferromagnets (FM), which could be important for magnetoelectric coupling, integrating ferroelectricity to control AM in multiferroic systems is largely missing. 

In this work, we reveal how many challenges to control AM can be overcome using (anti)ferroelectricity and a universal design principle illustrated in Fig.~1. We introduce an unexplored class of multiferroics: antiferroelectric altermagnets (AFEAM) which support electric control of spin in a single material using a small electric field and demonstrate exciting magnetoelectric coupling. 
Their opposite spin lattices are connected through rotation symmetry, $R$, which can also include a combination of translation or inversion, induced by the AFE environment to yield AM with spin polarization, $\textbf{\emph{P}}_S\neq0$. An applied electric field, $\varepsilon$, transforms AFEAM into conventional ferroelectric AFM (FEAFM) with $\textbf{\emph{P}}_S=0$, due to the direct translation, $t$, of the spin lattices in the FE environment. This transition enables electric control of $\textbf{\emph{P}}_S$. The mechanism and universality of our design principle are further demonstrated using an effective tight-binding (TB) model. To illustrate the practicality of our approach, we employ first-principles calculations to identify AFEAM in well-explored multiferroic materials, including (i) two-dimensional (2D) van der Waals (vdW) metal thio(seleno)phosphates~\cite{Susner2017:AM_MTPs} such as CuMoP$_2$S(Se)$_6$ and CuWP$_2$S(Se)$_6$, where AFEAM persists from monolayers (MLs) to bulk, independent of layer thickness and (ii) 3D multiferroic transition-metal-oxide perovskites~\cite{Si2024:antiferroelectric}, such as BiCrO$_3$.

Our design principle for the AFEAM and the electric control of $\textbf{\emph{P}}_S$ can be understood from the role of the AFE (FE) environment on AFM shown in Fig.~1, which leads to the AM (conventional AFM).
The influence of (anti)ferroelectricity on magnetism is analyzed using the exchange operation approach~\cite{noda2016:momentum,okugawa2018:weakly} and spin group theory~\cite{SpingroupPRX_Fang,SpingroupPRX_Song,SpingroupPRX_Liu,Liu2022:PRX,Qliu_spingroup_tool_arXiv}, employed to study momentum-dependent spin splitting in AM. The spin-dependent bands are described by the Kohn-Sham equations
\begin{equation}
\begin{aligned} 
\left[\frac12(\textbf{\emph{k}}-i\nabla)^2+V_\alpha\right]\psi_\alpha(\textbf{\emph{k}})=E_\alpha(\textbf{\emph{k}})\psi_\alpha(\textbf{\emph{k}}), 
\end{aligned} 
\label{eq:KS} 
\end{equation}
where \textbf{\emph{k}} is the wave vector, $V_\alpha$ is the Kohn-Sham potential with $\alpha$ for spin-up ($\uparrow$) and spin-down ($\downarrow$). $E$(\textbf{\emph{k}}) and $\psi$(\textbf{\emph{k}}) denote the eigenvalue and wave function. Exchange operations, $O$, are the symmetry operations in the geometric space group that exclude spin configurations. They map the spin-up to the spin-down lattices as $OV_\uparrow O^{-1}=V_\downarrow$. The spin degeneracy and splitting are determined
for $O\textbf{\emph{k}}=\textbf{\emph{k}}$, $E_\uparrow(\textbf{\emph{k}}) = E_\downarrow(\textbf{\emph{k}})$, and for $O\textbf{\emph{k}}\neq \textbf{\emph{k}}=\textbf{\emph{k}}^{\prime}$, $E_\uparrow(\textbf{\emph{k}})\neq E_\downarrow(\textbf{\emph{k}})=E_\downarrow(\textbf{\emph{k}}^{\prime})$~\cite{noda2016:momentum,okugawa2018:weakly}.

\begin{figure}[t]
\centering
\includegraphics*[width=0.47\textwidth]{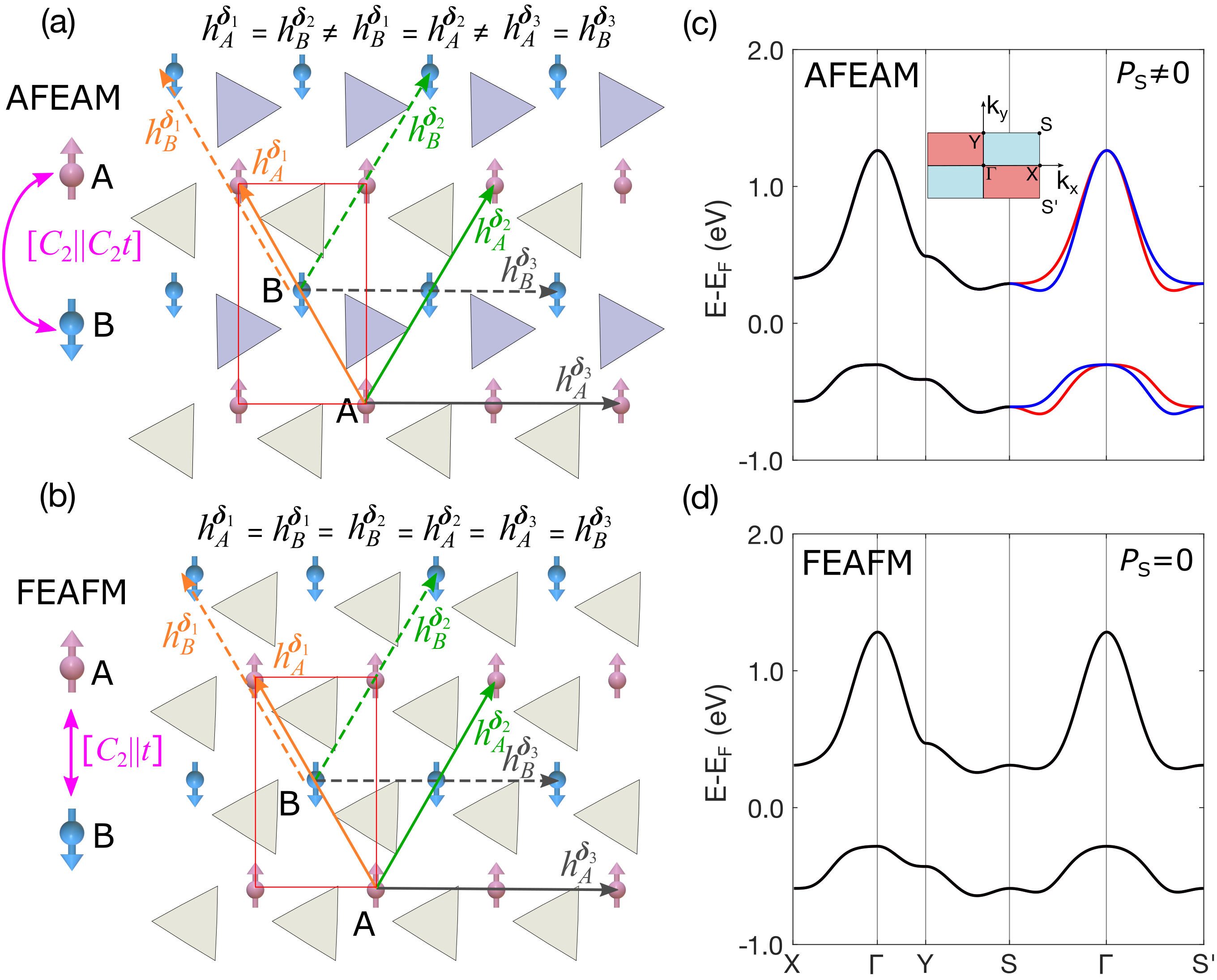}
\caption{Schematic of 2D rectangular  
magnetic sublattices A and B with AFE (distinct triangles) and FE (identical triangles) configurations, forming (a) AFEAM with $\left[C_2 \| C_2 t\right]$ symmetry and (b) FEAFM with $\left[C_2 \| t\right]$ symmetry, respectively. The solid (dashed) arrows indicate 3NN hopping vectors ($\boldsymbol{\delta}_{1-3}$) in A (B) sublattice with hopping strengths $h_A^{\boldsymbol{\delta}_{1-3}}$ ($h_B^{\boldsymbol{\delta}_{1-3}}$). The other three hopping vectors ($\boldsymbol{\delta}_{4-6}=-\boldsymbol{\delta}_{1-3}$) are not shown as the hopping strengths along opposite directions are identical ($h_{A, B}^{\boldsymbol{\delta}_{4-6}}=h_{A, B}^{\boldsymbol{\delta}_{1-3}}$). (c) Calculated bands for (a) AFEAM using TB model with parameters $f_{A, B}^{\eta_{1-6}}$ = 0.10 eV, $g_{A, B}^{\kappa_{1-6}}$ = 0.06 eV, $h_A^{\boldsymbol{\delta}_{1,4}}=h_B^{\boldsymbol{\delta}_{2,5}}$ = 0.04 eV, $h_A^{\boldsymbol{\delta}_{2,5}}=h_B^{\boldsymbol{\delta}_{1,4}}$= 0.01 eV, $h_A^{\boldsymbol{\delta}_{3,6}}=h_B^{\boldsymbol{\delta}_{3,6}}$= 0.03 eV, and $M_A=-M_B$ = 0.45 eV. The black, red, and blue denote the spin-degenerate, -up, and -down bands, respectively. The inset diagram indicates the $\textbf{\emph{P}}_S$ sign in the Brillouin zone of (a). (d) Same as (c), but for (b) FEAFM with $h_A^{\boldsymbol{\delta}_{1-6}}=h_B^{\boldsymbol{\delta}_{1-6}}$ = 0.03 eV.}
\label{fig:F2}
\end{figure} 

For FEAFM, with net electric polarization, $\textbf{\emph{P}}\neq0$, spin lattices are related by direct $t$. Since $t$\textbf{\emph{k}} = \textbf{\emph{k}} for all \textbf{\emph{k}}, the bands are spin-degenerate in the whole Brillouin zone, giving the conventional AFM with $\textbf{\emph{P}}_S=0$ as in Fig.~1(b). In contrast, for AFEAM, with alternating electric polarization (\textbf{\emph{P}}=0), its spin lattices cannot be connected by a direct $t$, but 
by $R$ (commonly $C_2t$), leading to $\textbf{\emph{P}}_S=0$ only along specific paths where $R\textbf{\emph{k}} = \textbf{\emph{k}}$, while $\textbf{\emph{P}}_S\neq0$ elsewhere, defining AM as in Fig.~1(a). 

Considering AFE and FE configurations are locked with AM and AFM states and their transition can be flexibly reversed through $\varepsilon$ ~\cite{Si2024:antiferroelectric,xue2022_AFEReview_Matter}, we propose $\varepsilon$-controlled AFEAM and FEAFM transitions to toggle the $\textbf{\emph{P}}_S$ as shown in Fig.~1. This electric control of $\textbf{\emph{P}}_S$ relying on straightforward conditions, where the symmetry of AFM can be switched between $\left[C_2 \| t\right]$ and $\left[C_2 \| R\right]$, offers a novel and accessible mechanism for realizing magnetoelectric coupling.

\begin{figure*}[t]
\centering
\includegraphics*[width=0.8\textwidth]{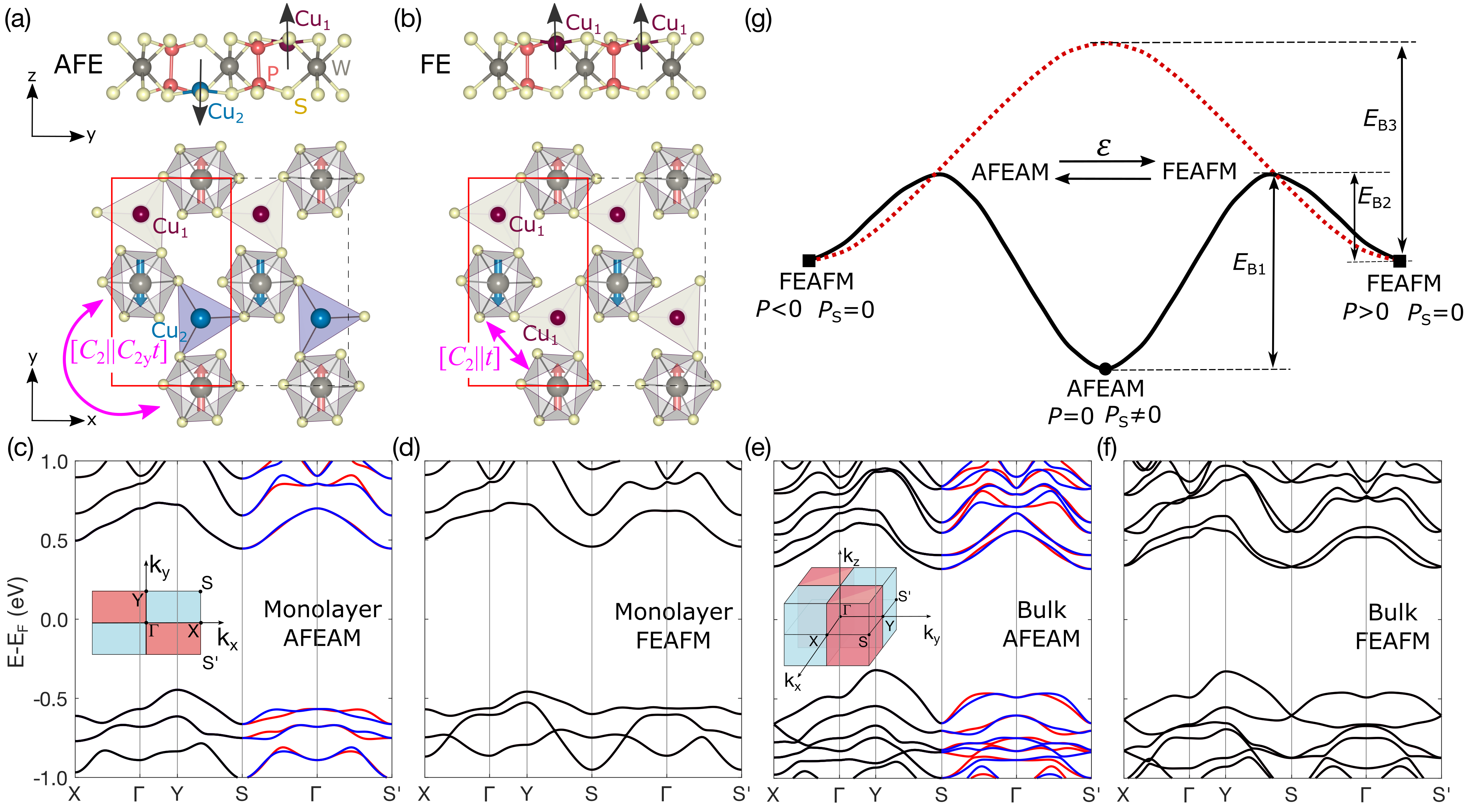}
\caption{(a) and (b) Structures and magnetic ground state (sAFM) of ML AFE and FE CuWP$_2$S$_6$, characterized by $\left[C_2 \| C_2 t\right]$ and by $\left[C_2 \| t\right]$ symmetry, respectively. (c) and (d) Calculated bands for (a) and (b), where the black, red, and blue denote the spin-degenerate, -up, and -down bands, respectively. (e) and (f) Same as (c) and (d) but for bulk CuWP$_2$S$_6$. The insets in (c) and (e) show $P_S$ sign in their Brillouin zones. (g) Calculated kinetic switching pathways between negative ($\textbf{\emph{P}}<0$) and positive ($\textbf{\emph{P}}>0$) FE states through AFE (solid line) or PE (dotted line) states in ML CuWP$_2$S$_6$. $E_{B1}$, $E_{B2}$, and $E_{B3}$ indicate the energy barriers of AFE-to-FE, FE-to-AFE, and FE-to-PE-to-FE states, respectively. $\textbf{\emph{P}}_S$ can be electrically toggled by switching between AFEAM and FEAFM.} 
\label{fig:F3}
\end{figure*}

To illustrate our design principle, we develop a TB model based on a general 2D rectangular lattice with nested AFM order (Fig.~2). This setup includes the essential elements of our proposal: AFM sublattices with AFE (FE) configurations connected by $C_{2}t$ ($t$) symmetry. The minimal model, to capture the influence of AFE and FE configurations on AM, incorporates up to the third-nearest-neighbor hopping 
\begin{equation}
\begin{aligned}
H & =\left(\sum_{i, j}\left(f_i^{\eta_j} c_i^{\dagger} c_{i+\eta_j}+g_i^{\kappa_j} c_i^{\dagger} c_{i+\kappa_j}+h_i^{\delta_j} c_i^{\dagger} c_{i+\delta_j}\right)+\text { h.c. }\right) \\
& +M_{A, B} \sum_{i \in A, B} c_i^{\dagger} \sigma_z c_i. 
\end{aligned}
\end{equation}
Here, $c_i^{\dagger}\left(c_i\right)$ are electron creation (annihilation) operators at site $i$ and $\sigma$ is the Pauli matrix. The first three terms describe hoppings between first (NN), second (2NN), and third (3NN) nearest-neighbors with hopping parameters $f_i^{\boldsymbol{\eta}_j}$, $g_i^{\boldsymbol{\kappa}_j}$, and $h_i^{\boldsymbol{\delta}_j}$, where $\boldsymbol{\eta_j}$, $\boldsymbol{\kappa}_j$, and $\boldsymbol{\delta}_j$ are the vectors connecting site $i$ to its NN, 2NN, and 3NN sites. The last term indicates AFM exchange fields on sublattices A and B with $M_A$=-$M_B$.

The emergence of AM arises from the inequivalence between magnetic sublattices~\cite{Vsmejkal2022:emerging}. While the NN and 2NN hoppings do not contribute to AM as they are sublattice-equivalent under both AFE and FE configurations, as discussed in Supplemental Material (SM)~\cite{SMAFEAM}, 3NN hoppings are dependent on sublattices and (A)FE configurations, playing a crucial role in altermagnetism. In FEAFM, $\left[C_2\|t\right]$ symmetry ensures that 3NN hoppings are also equivalent between opposite sublattices ($h_A^{\boldsymbol{\delta}_{1-6}}=h_B^{\boldsymbol{\delta}_{1-6}}$)[Fig.~2(b)]. Consequently, the calculated bands exhibit spin degeneracy, characteristic of conventional AFM [Fig.~2(d)]. In contrast, in AFEAM, the AFE configuration breaks $t$ symmetry, leaving magnetic sublattices connected by $\left[C_2\|C_2t\right]$. This symmetry breaking alters the 3NN hopping strengths such that hopping along $\delta_{1,4}$ in sublattice A ($h_A^{\boldsymbol{\delta}_{1,4}}$) differs from that in sublattice B ($h_B^{\boldsymbol{\delta}_{1,4}}$) but equals the hopping along $\delta_{2,5}$ ($h_A^{\boldsymbol{\delta}_{2,5}}$) in sublattice B, and vice versa, i.e. $h_A^{\boldsymbol{\delta}_{1,4}}=h_B^{\boldsymbol{\delta}_{2,5}} \neq h_B^{\boldsymbol{\delta}_{1,4}}=h_A^{\boldsymbol{\delta}_{2,5}} \neq h_A^{\boldsymbol{\delta}_{3,6}}=h_B^{\boldsymbol{\delta}_{3,6}}$ as depicted in Fig.~2(a). This directional alternation induces AM, as confirmed by the calculated bands in Fig.~2(c). This effective model, clearly demonstrating our AFEAM design, also provides a versatile framework for exploring AM mechanisms and applications.

Following the above TB model, we first examine our design principle for AFEAM and $\varepsilon$-controlled $\textbf{\emph{P}}_S$ in the family Cu$M$P$_2$S(Se)$_6$, where $M$ = Cr, Mo, W, known for its 2D multiferroic properties~\cite{Qi2018:APL,Hua2021:CPL_CWPS,io2023direct,wang2023electrical,Hu2024:CCPS_exp,yu2023voltage}. 
The Cu ions have a zigzag vertical shift from the mid-plane of its structure, inducing a local out-of-plane electric polarization
with AFE order, while the Cr/Mo/W atoms induce magnetism constructing spin lattices as in Fig.~3(a). Applying $\varepsilon$ can align the Cu ions to the same side within the vertical layer, producing a FE phase [Fig.~3(b)], as observed in recent experiments~\cite{io2023direct,wang2023electrical,Hu2024:CCPS_exp}. 

To reveal the magnetic order and electronic structure of Cu$M$P$_2$S(Se)$_6$, we first explore their MLs and then study thicker films and bulk material through vdW stacking. We consider three typical magnetic configurations: FM, stripe AFM (sAFM) in Fig.~3, and zigzag AFM in SM {~\cite{SMAFEAM}}. Our calculations show that the ML CuCrP$_2$S(Se)$_6$ favors FM, while MLs of CuMoP$_2$S(Se)$_6$ and CuWP$_2$S(Se)$_6$ favor sAFM as discussed in SM~\cite{SMAFEAM}, consistent with the previous studies and experiments~\cite{Qi2018:APL,Hua2021:CPL_CWPS,io2023direct,wang2023electrical,Hu2024:CCPS_exp}. This $M$-type dependency is attributed to the combined 
effect~\cite{Hua2021:CPL_CWPS,Duan2022:TMP2X6,Huang2018:jacs} of the crystal field splitting and the exchange interaction~\cite{SMAFEAM}.
While previous studies have mainly focused on FM CuCrP$_2$S(Se)$_6$, the AFM CuMoP$_2$S(Se)$_6$ and CuWP$_2$S(Se)$_6$ are largely overlooked as AFM are often deemed less interesting. However, these AFM are excellent candidates for our AFEAM design in Fig.~1.

Without losing generality, we consider ML CuWP$_2$S$_6$ (CWPS) to demonstrate our AFEAM design. Additional results for other Cu$M$P$_2$S(Se)$_6$ compounds are provided in SM~\cite{SMAFEAM}. With ignoring the magnetism, the space group of AFE ML CWPS is $P2_1$ with symmetry $\{E, C_{2y}t\}$, for the identity and screw axis along $y$. Instead, considering the ground-state sAFM order, its space group is lowered to $P_1$ containing only $\{E\}$ with missing $C_{2y}t$ acting as the key exchange operations that define the electronic structures through the spin group [$C_2 \Vert C_{2y}t$], as shown in Fig.~3(a). 

The exchange operation of $C_{2y}t$ satisfies $C_{2y}t\textbf{\emph{k}}(\Gamma–\text{Y}) = \textbf{\emph{k}}(\Gamma–\text{Y})$, where $\textbf{\emph{k}}(\Gamma–\text{Y})$ indicates the $\textbf{\emph{k}}$ points along the $\Gamma$–Y path [Fig.~3(c)]. Moreover, considering inversion $I$, we can get an additional exchange operation, $IC_{2y}t$, which satisfies 
$IC_{2 y}t\mathrm{~V}_{\uparrow}\left(IC_{2y} t\right)^{-1}=V_{\downarrow}$ and $IC_{2y}t\textbf{\emph{k}}(\Gamma–\text{X}) = \textbf{\emph{k}}(\Gamma–\text{X})$. Thus, we obtain $E$$_\uparrow$(X$-\Gamma–$Y)  = $E$$_\downarrow$(X$-\Gamma–$Y), giving the spin degeneracy along X$-\Gamma–$Y. In contrast, for other paths lacking specific symmetry operations, such as S–$\Gamma$–S$^\prime$, their bands are spin polarized. Furthermore, the relation $C_{2y}t$\textbf{\emph{k}}($\Gamma$–S) = \textbf{\emph{k}}($\Gamma$–{S}$^\prime$) introduces $E$$_\uparrow$($\Gamma$–S) = $E$$_\downarrow$($\Gamma$–{S}$^\prime$), resulting in opposite $\textbf{\emph{P}}_S$ along such two paths with a $d$-wave character [Fig.~3(c)]. The spin splitting in CWPS is up to 120 meV, which can be experimentally distinguished by the angle-resolved photoemission spectroscopy~\cite{zhu2024:MnTe2,Krempasky2024:MnTe}. 
Since, unlike the spin-orbit coupling (SOC), AM breaks the 
time-reversal symmetry and can display 
typical ferromagnetic properties, we can use spin transport to distinguish altermagnetic and SOC spin splitting, as discussed in SM~\cite{SMAFEAM}. For FE-sAFM CWPS, there is a direct $t$ operation as $O$ [Fig.~3(b)] satisfying $t\textbf{\emph{k}}$ = $\textbf{\emph{k}}$ for all \textbf{\emph{k}}, thus there is no spin splitting throughout the Brillouin zone, giving a conventional AFM [Fig.~3(d)]. This phenomenon, where AFE states produce AM, while FE states revert to conventional AFM, illustrates our design principle and confirms CWPS as AFEAM. 

Considering CWPS is a vdW material, its properties of the bulk structure and few-layer films~\cite{SMAFEAM} are expected to resemble those of the MLs. This expectation is supported by our first-principles calculations, which confirm both the bulk and film forms of CWPS retain the same intralayer AFE-sAFM order and show AFEAM behavior, as in Fig.~3(e) and SM~\cite{SMAFEAM}. We find the sign of the $\textbf{\emph{P}}_S$ in the bulk CWPS does not vary with $k_z$, exhibiting altermangnetic spin splitting for all $k_z$ planes (even for $k_z$ = 0). This behavior, distinct from the reported bulk AM ~\cite{Krempasky2024:MnTe,Osumi2024:PRB_MnTe,Lee2024:PRL_MnTe,Reimers2024:NC_CrSb,ding2024large:CrSb,zeng2024observation:CrSb,yang2024three:CrSb}, reflects the vdW nature of CWPS. Furthermore, by altering the AFE to FE order, bulk CWPS also switches from AFEAM to FEAFM [Fig.~3(f)]. For CWPS films, our calculations confirm their electronic structures are consistent with those of their ML and bulk counterparts~\cite{SMAFEAM}. Such properties, independent of the layer thickness, facilitate the experimental realization of our proposal.

The AFE-to-FE transition typically occurs reversibly when the applied $\varepsilon$, with the energy $E_\varepsilon$,
 aligns the electric polarization by overcoming its energy barrier, $E_{B1}$, while the FE-to-AFE transition depends on the stability of the FE state~\cite{Si2024:antiferroelectric,xue2022_AFEReview_Matter}. An unstable FE state spontaneously reverts to the AFE ground state when the $\varepsilon$ is removed~\cite{Si2024:antiferroelectric}, while a metastable FE state requires a reversed $\varepsilon$ to overcome the metastability barrier, $E_{B2}$ and return to the AFE state~\cite{xue2022_AFEReview_Matter,Duan_ACSNano_AFE_GeSe,Balke_ACSNano_AFE_CIPS}. To examine these transitions in ML CWPS, we calculate the kinetic switching pathways using the climbing image nudged elastic band method, a widely proven tool for modeling such transitions~\cite{Duan_ACSNano_AFE_GeSe}. Our results [Fig.~3(g)] show that ML CWPS has an AFE ground state and a metastable FE state. The AFE-to-FE transition occurs when $E_{\varepsilon} > E_{B1}$, while the FE-to-AFE transition is driven by a reversed $\varepsilon$ with $E_{\varepsilon} > E_{B2}$. To avoid switching the FE state to its opposite $\textbf{\emph{P}}$, the $\varepsilon$ should satisfy $E_{B2} < E_{\varepsilon} < E_{B1,3}$, where $E_{B3}$ represents the energy barrier through the paraelectric (PE) state. The calculated $E_{B1}$ ($E_{B2}$, $E_{B3}$) in ML CWPS is 0.33 (0.15, 0.37) eV/f.u., comparable with similar transitions in CuInP$_2$S$_6$ and CuCrP$_2$S$_6$, where a weak $\varepsilon$ (tens of kV/cm) can experimentally drive the AFE-FE transition~\cite{Balke_ACSNano_AFE_CIPS,wang2023electrical}. Given the structural and polarization similarities among these materials, we estimate that an $\varepsilon$ of similar magnitude should be sufficient to induce phase transitions in the CWPS system. This highlights the potential of CWPS for $\varepsilon$ control of $\textbf{\emph{P}}$ and $\textbf{\emph{P}}_S$.

\begin{figure}[t]
\centering
\includegraphics[width=0.49\textwidth]{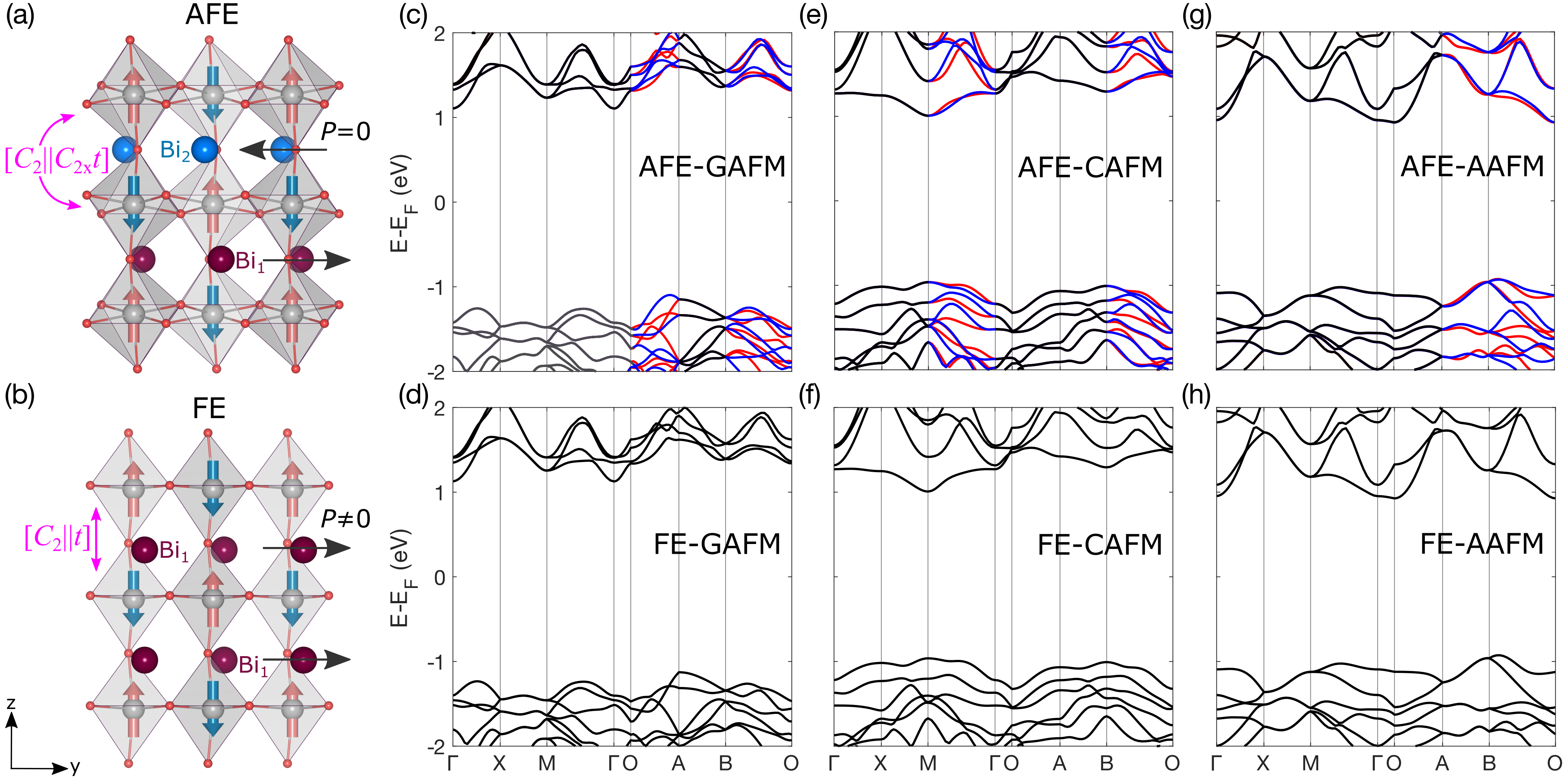}
\caption{(a) Structure of BiCrO$_3$ showing AFE-GAFM ground state with their spin lattices connected by $C_{2x}t$. 
(b) As in (a) but showing FE-GAFM state with spin lattices connected by $t$. 
(c)-(d) Spin-resolved bands for (a)-(b), respectively. (e)-(f) and (g)-(h) As in (c)-(d), but for CAFM, and AAFM states, respectively.}
\label{fig:F4}
\end{figure}

Expanding our focus to other multiferroics, we explore 3D oxide perovskites, highlighting the broad applicability of our design principle across various material dimensions and types. Traditional AFE perovskites, such as PbZrO$_3$, PbHfO$_3$, and NaNbO$_3$, have been studied for their applications in energy storage, sensors, and memory devices~\cite{Si2024:antiferroelectric}. By including magnetic atoms, multiferroic perovskites, as exemplified with BiFeO$_3$, have been extensively studied~\cite{yang2015:bifeo3,farooq2023:spontaneous}. We consider a multiferroic BiCrO$_3$ to demonstrate our design principle due to its known coexistence of AFE and AFM properties~\cite{hill2002:first,kim2006:antiferroelectricity}. In its ground state, as shown in Fig.~4(a), BiCrO$_3$ is a G-type AFM (GAFM) and belongs to the $Pbnm$ space group. Its AFE behavior results from the opposing displacements of the adjacent Bi atoms. $\varepsilon$ can shift the Bi atoms to the same side~\cite{kim2006:antiferroelectricity}, changing the system from AFE to FE states as in Fig.~4(b). Our calculations show that its ground state of AFE-GAFM gives AFEAM, characterized by the [$C_2 \Vert C_{2x}t$], exhibiting altermagnetic spin splitting [Fig.~4(c)]. For FE-GAFM case, its spin lattices are directly connected by [$C_2 \Vert t$], showing FEAFM [Fig.~4(d)]. Such phenomenon in BiCrO$_3$ further corroborates our design principle in Fig.~1. 

While BiCrO$_3$ predominantly exhibits GAFM, we have also explored the common C-type (CAFM) and A-type (AAFM) orders of AFM in perovskites~\cite{SMAFEAM}. Similar to AFE-GAFM, both AFE-CAFM and AFE-AAFM BiCrO$_3$ display AM with spin splitting along distinct paths [Figs.~4(e) and 4(g)], but this splitting disappears in their FE phases [Figs.~4(f) and 4(h)]. Our findings on such altermagnetic behavior in BiCrO$_3$ can be directly extended to other similar perovskites by substituting its Bi or Cr elements~\cite{ACrO3_2017PRB,BiMO3_2010PRB,SMAFEAM}, provided that the materials retain the AFE and AFM configurations. The abundance of such perovskites enriches the scope for further experimental validation of how altermagnetic spin splitting is affected by magnetic orders and crystalline directions.

Our research establishes a universal design principle for AFEAM, positioning them as integral components of multiferroic materials. This design provides a versatile platform to explore magnetoelectric coupling influenced by the interplay between ferroelectricity and magnetism. The ability to electrically control the  
transition between AFEAM ( $\textbf{\emph{P}}=0$, $\textbf{\emph{P}}_S\neq0$) and FEAFM ($\textbf{\emph{P}}\neq0$, $\textbf{\emph{P}}_S=0$) with  
a small $\varepsilon$, not only opens new avenues for investigating electrically controlled spintronics but also enhances their potential in multiferroics, particularly in tunnel junctions~\cite{Shao2024:npjspintronics}. While there is a debate about the terminology and the origin of the concepts of AM and nonrelativistic spin splitting, studied for decades~\cite{Pekar1964:JETP}, our findings are not limited to that debate as they offer unexplored opportunities for magnetoelectric coupling and novel classes of multiferroics, independent of the specific terminological details and if the SOC will invalidate a strict AM definition. Notably, unlike spin torque to electrically control magnetism~\cite{Yang2022:N,Tsymbal:2019}, our electric control does not require magnetization reversal and can be realized using a small $\varepsilon$, thereby enhancing both the speed and the energy efficiency of potential devices.

Moreover, just as the discovery of 2D FM have advanced opportunities for designing vdW heterostructures, with the prospect of a growing family of 2D AM, it would be important to understand how they can transform the neighboring materials through proximity effects~\cite{Zutic2019:MT,Ibrahim2020:2DM,Wang2024stackingAM_PRB,Guo2024valleyAM_PRB}. Since electric control of topological properties is already considered in various heterostructures~\cite{chen2024ferroelectricity,Zhou2021:PRL}, we expect that our proposal for electrically controlled vdW multiferroics, with multiple broken symmetries, can further expand the family of topological states and facilitate their manipulation, not only in the normal~\cite{Guo2023:quantum,Fernandes2024:PRB}, but also in superconducting states~\cite{AMMajorana_PRL2024,Li2023:PRB,zhou2022fusion,Amundsen2024:RMP}.
 
\begin{acknowledgements} We acknowledge Cheng Zhang, Yuan Lu, and Cheng Song for valuable discussions about their related experimental efforts. Tong Zhou thanks Avery Zhou and Mila Zhou for their inspiration. This work is supported by the National Natural Science Foundation of China (Grants No. 12474155, 12447163, 12488101, and 11904250), the Innovation Program for Quantum Science and Technology (Grant No. 2021ZD0302800), and the Air Force Office of Scientific Research under Award No. FA9550-22-1-0349 (Y.L., I.\v{Z}.). The computational resources for this research were provided by the High Performance Computing Platform at the Eastern Institute of Technology, Ningbo. 

X. Duan and J. Zhang contributed equally to this work.

\end{acknowledgements} 
\bibliography{AFEAM_bib_SMRefs}
\end{document}